# Single-substrate Enzyme Kinetics: The Quasi-steady-state Approximation and Beyond

Sharmistha Dhatt[#] and Kamal Bhattacharyya[*]

Department of Chemistry, University of Calcutta, Kolkata 700 009, India

Abstract

We analyze the standard model of enzyme-catalyzed reactions at various substrate-enzyme ratios by adopting a different scaling scheme. The regions of validity of the quasi-steady-state approximation are noted. Certain prevalent conditions are checked and compared against the actual findings. Efficacies of a few other measures are highlighted. Some recent observations are rationalized, particularly at moderate and high enzyme concentrations.

Keywords: Michaelis-Menten kinetics, Quasi-steady state approximation, Padé approximants

[#]pcsdhatt@gmail.com

[*]pchemkb@gmail.com (Corresponding author)

## 1. Introduction

The kinetics of enzyme-catalyzed reactions is usually modeled by a set of coupled differential equations, exact solutions of which are not obtainable in closed forms. However, in view of the importance of such reactions involving biochemical systems, simplifying assumptions are often made. One celebrated result of such endeavors is the standard Michaelis–Menten (MM) form, based on the quasi-steady-state approximation (QSSA). Crucial in this treatment is the assumption that, after a short transient, the concentration of the enzyme-substrate complex remains approximately constant. The impact of the MM form is still quite significant, even after about a century, as may be gazed from a very recent attention to the original work [1].

The two-step model is symbolized by the reaction scheme

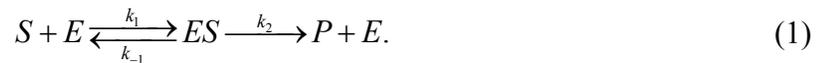

$$S + E \underset{k_{-1}}{\overset{k_1}{\rightleftarrows}} ES \xrightarrow{k_2} P + E. \tag{1}$$

On the basis of (1), various aspects of the QSSA have been studied from time to time. An early study by Laidler [2] revealed that any of the following four conditions is necessary for the applicability of the QSSA:

$$\begin{aligned}&(a)\ s_0/e_0 \gg 1;\\&(b)\ s_0/e_0 \ll 1;\\&(c)\ k_1 s_0/(k_{-1}+k_2) \ll 1;\\&(d)\ k_1 e_0/(k_{-1}+k_2) \ll 1.\end{aligned} \tag{2}$$

In (2), $s_0$ and $e_0$ refer to the initial concentrations of substrate and enzyme, respectively. Later, Laidler *et al* [3] also noted that the product profile gives a clear signature of the validity of QSSA. Most authors [4 - 11], however, opine in favor of condition (2a) only. A somewhat different condition for the validity of QSSA has also been suggested by some authors [12, 13]. It reads as

$$e_0/(s_0 + K_m) \ll 1, \tag{3}$$

where $K_m$ is the Michaelis constant, defined by

$$K_m = (k_{-1} + k_2)/k_1. \tag{4}$$

Borghans *et al* [14] distinguished different types of QSSA as standard QSSA (s-QSSA), reverse QSSA (r-QSSA) and total QSSA (t-QSSA), depending on whether the ratio $s_0/e_0$ is large, small or arbitrary. Several works [15 - 17] then concentrated on moderate to high enzyme-substrate ratios. A result of Tzafriri *et al* [17] is notable in this context. They remarked that condition (3) retains its validity only when $s_0/e_0$ is large. In case $e_0/s_0$ is large, it should be modified as

$$s_0 / (e_0 + K_m) \ll 1. \tag{5}$$

Analyses over a wide range of the ratio $s_0/e_0$ were pursued in several recent theoretical works [18 - 21] with interesting experimental relevance [22]. A route to calculate rate constants also followed [23].

A few problems, however, remained unsolved. Thus, following an earlier work [17], Kargi [20] mainly focused attention on the ratio $s_0/e_0$ at both ends and concluded again that QSSA can be implemented in either case, including the intermediate region. A very recent work [21], on the other hand, maintained that QSSA is valid, if at all, only for a short time-interval when $e_0/s_0$ is large and that the region of validity is considerably larger for large $s_0/e_0$.

In view of the above remarks, our purpose is to explore whether (i) QSSA is valid irrespective of the value of the ratio $s_0/e_0$, (ii) the steady-state region, if exists, is smaller for large $e_0/s_0$ ratio, (iii) the initial transient period is small when QSSA holds, (iv) the set of conditions (2), or (3) or (5) works sensibly in predicting the legitimacy of the QSSA, and (v) better measures of the applicability of QSSA exist. Answers to such questions can hopefully settle the issue of applicability of QSSA for moderate-to-large ratio of $e_0/s_0$ [14 - 20] or only small $e_0/s_0$ [4 – 13, 21].

## 2. The method

On the basis of (1), the following differential equations emerge:

$$\frac{d[S]}{dt} = -k_1[E][S] + k_{-1}[ES], \tag{6}$$

$$\frac{d[ES]}{dt} = k_1[E][S] - k_{-1}[ES] - k_2[ES], \tag{7}$$

$$\frac{d[E]}{dt} = -k_1[E][S] + k_{-1}[ES] + k_2[ES], \tag{8}$$

$$\frac{d[P]}{dt} = k_2[ES]. \tag{9}$$

In addition, we have two mass conservation equations

$$e_0 = [E] + [ES], \tag{10a}$$

$$s_0 = [S]+[ES]+[P]. \tag{10b}$$

To solve the above equations, we employ the following set of new dimensionless variables:

$$\alpha = \frac{[E]}{e_0}, \beta = \frac{[S]}{e_0}, \gamma = \frac{[ES]}{e_0}, \delta = \frac{[P]}{e_0}, \tau = k_2 t. \tag{11}$$

Then, relevant kinetic equations, out of (6) - (9), may be compactly written as

$$\frac{d\alpha}{d\tau} = \frac{d\beta}{d\tau} + (1-\alpha), \tag{12}$$

$$\frac{d\beta}{d\tau} = -K_1 \beta \alpha + K_2 (1-\alpha), \tag{13}$$

with initial conditions

$$\alpha_0 = 1, \beta_0 = s_0/e_0. \tag{14}$$

The constants $K_1$ and $K_2$ in (13) are given by

$$K_1 = k_1 e_0/k_2, K_2 = k_{-1}/k_2. \tag{15}$$

They will play a key role in predicting the validity of QSSA in a particular context, as we shall soon see. The conservation equations (10) read now as

$$\alpha + \gamma = 1, \beta + \gamma + \delta = \beta_0. \tag{16}$$

The above system of non-linear equations (12) – (13), with the aid of (16), can be solved analytically using the traditional power series method. Expressing the concentrations of the participating species in power series of $\tau$, *viz.*,

$$\alpha_\tau = \sum_{j=0} \alpha_j \tau^j, \beta_\tau = \sum_{j=0} \beta_j \tau^j, \gamma_\tau = \sum_{j=0} \gamma_j \tau^j, \delta_\tau = \sum_{j=0} \delta_j \tau^j, \tag{17}$$

inserting them suitably into (12) and (13), and collecting similar powers of $\tau$, the unknown parameters of the expansions are obtained. Note that our scaling is different from others [15, 24], but probably more useful. Moreover, to tackle the expansions in (17) at large $\tau$, we construct three types of Padé approximants [25], [N/N], [(N+1)/N] and [N/(N+1)]. This Padé method has been found to be quite faithful in such and varied contexts [26 - 33]. It is tested to be faithful here too.

Numerical stability of our computations is checked via two routes. First, the estimates are considered reliable only when at least two varieties of the above Padé sequences agree. Secondly, one finds from (12), (13) and (16) that

$$d\gamma/d\tau = K_1 \beta - (K_1 \beta + K_2 + 1)\gamma. \tag{18}$$

Therefore, the point $\tau_c$ at which $\gamma$ attains its maximum value ($\gamma_c$) is obtainable from (18) as

$$\gamma_c = K_1\beta_c/(K_1\beta_c + K_2 + 1). \tag{19}$$

Hence, from the temporal profiles of the complex and substrate concentrations, we are in a position to verify how far the left hand part agrees with the right side. We have also noted that, by following this procedure, one can go well beyond the region of adequacy of QSSA, and hence can assess the quality of the steady state, if there is any.

## 3. Legitimacy of QSSA

Bajer *et al* [21] considered different cases and noted the concentration profiles of the components. So, let us concentrate on the $\gamma$-$\tau$ plot first.

It follows from (18) that

$$(d\gamma/d\tau)_{\tau_2} - (d\gamma/d\tau)_{\tau_1} = K_1\left(\beta_{\tau_2} - \beta_{\tau_1}\right) - K_1\left(\beta_{\tau_2}\gamma_{\tau_2} - \beta_{\tau_1}\gamma_{\tau_1}\right) - (K_2+1)(\gamma_{\tau_2} - \gamma_{\tau_1}). \tag{20}$$

For an observable region over which QSSA will be valid, one needs to consider $\tau_2 > \tau_1 > \tau_c$. It is also apparent that, if such a region exists up to $\tau_2$, one would have

$$(d\gamma/d\tau)_{\tau_2} \approx (d\gamma/d\tau)_{\tau_1} \approx 0;\ \gamma_{\tau_2} \approx \gamma_{\tau_1}. \tag{21}$$

Condition (21) is satisfied by (20) only if

$$K_1 \ll 1. \tag{22}$$

Thus, (22) turns out to be one important necessary condition for the satisfaction of QSSA.

Another deductive analysis to justify the QSSA arises out of the observation that $\beta_\tau$ shows a linear fall-off beyond $\tau = \tau_c$ up to the range of validity of the approximation. Thus, we can write

$$\beta_\tau = \beta_c + \overline{\beta}(\tau - \tau_c),\ \tau > \tau_c. \tag{23}$$

From (12), however, one notes that

$$(d\beta/d\tau)_{\tau=\tau_c} = -\gamma_c. \tag{24}$$

Therefore, one obtains from (23) and (24) that

$$\overline{\beta} = -\gamma_c. \tag{25}$$

We now pay attention to (13) and write

$$-\gamma_c \approx -K_1\beta_\tau(1-\gamma_\tau) + K_2\gamma_\tau. \tag{26}$$

Putting (23) and (25) in (26), a rearrangement leads to

$$\gamma_\tau \approx \gamma_c + (\gamma_c - 1)\frac{K_1\gamma_c}{K_1\beta_c + K_2}(\tau - \tau_c), \tag{27}$$

correct up to first order in ($\tau - \tau_c$). The point now is that, if $\gamma$ has to show a maximum at $\tau = \tau_c$, then we expect no first-order term in the expansion (27). This implies, the coefficient associated with this term should vanish. In other words, we have the condition

$$|K_1\gamma_c/(K_1\beta_c + K_2)| << 1. \tag{28}$$

Since $\beta_c$ or $\gamma_c$ can vary arbitrarily, a sufficient condition for the satisfaction of (28) is

$$K_2/K_1 >> 1. \tag{29}$$

In effect, therefore, we have arrived at two conditions here for the validity of QSSA. The first one (22) is a necessary condition, while the second one (29) is sufficient.

**Results and discussion**

We are now in a position to explore the adequacy of QSSA. Table 1 shows sample results with same $s_0/e_0$ as chosen by Bajer *et al* [21]. We also report here the $\tau_c$ value in each case under study. For a fixed ratio of $s_0/e_0$, we choose to demonstrate here two distinct situations. The first entries do not conform to QSSA, while the second ones do. The corresponding concentration profiles for the complex are shown in Figures 1 to 3. It is evident from the figures that cases (*a*) refer to those where QSSA is not obeyed. Cases (*b*), however, reveal profiles in tune with QSSA. Interesting here is the set with $s_0/e_0 = 1/10$, as Figure 3 shows. This case is chosen with $K_2/K_1 = 10$ only. Had the value been larger, one would find a still longer steady region. Indeed, cases (*b*) of Figures 2 and 3 support Kargi's [20] assertions.

Table 1 displays, in addition, the various criteria used by several authors from time to time. They include conditions 2(*a*) – 2(*d*), (3) and (5). Of these, (3) is most popular. However, satisfaction of a particular condition in the form displayed earlier is only qualitative in character. So, we have fixed here a standard. If a quantity '*q*' is said to obey $q << 1$, we accept a value of 1/10 or lower. Similarly, we allow $q \geq 10$ to imply $q >> 1$. In Table 1, satisfaction of a condition is reported by 'yes/no' type response. One immediately notes that condition (3) does not reveal the truth at the very first three rows. Since condition (5) is said to work only for large $e_0$, we show its worth only at appropriate places. Needless to mention, it does not come up as a right criterion too. Even, one cannot go by majority. The first three rows again bear clear signatures in this respect. As an alternative, if we stick to our criteria of low $K_1$ and high $K_2/K_1$, the observations quoted in the table can all be rationalized. Let us also notice that a small $\tau_c$ (or $t_c$) is not a necessary condition for QSSA.

Table 2 reveals clearly the fates of the sets chosen by Bajer et al [21]. On the basis of our criteria, we expect QSSA to hold in their sets A and B if $s_0/e_0 \geq 1$. Definitely, in the excess enzyme case, they had chosen the wrong set G to observe a steady state. If at all, QSSA might have been true in this situation for set B. Better, they should have chosen a different set of values for ($k_1$, $k_{-1}$, $k_2$) like (0.1, 10, 10), or still better (0.1, 100, 100). One is also not sure about the adequacy of QSSA for set G in their $s_0/e_0 = 10$ case either. Of course, a large $\beta_c$ in (28) can somehow favor the situation, but their figure with an insensitive ordinate for $\gamma_\tau$ seems inconclusive with respect to attainment of the steady state.

In brief, we have thus found that QSSA (i) can be acceptable at both moderate and large $e_0/s_0$ under conditions similar to those applicable to large $s_0/e_0$, (ii) can hold over a good span of time at large $K_2/K_1$, (iii) can show up even when $\gamma_c$ or $\tau_c$ is not small enough, and (iv) requires validity of conditions (22) and (29). We also possibly have settled the issues raised by some recent works [20, 21]. Thus, MM kinetics can hold irrespective of the enzyme-substrate ratio. In fine, our work examines the kinetic equation without any restriction on the enzyme-substrate ratio to encompass both classical *in-vitro* as well as biotechnologically-pro *in-vivo* situations.

**Acknowledgement**

Table 1. Characteristics of sets with varying ratios of $s_0/e_0$ and $K_2/K_1$ and at several $K_1$ values. The observations correspond to Figures 1 – 3.

| $s_0, e_0$ | $\tau_c$ | $(k_1, k_{-1}, k_2)$ | Conditions 2(a) | 2(b) | 2(c) | 2(d) | 3 | 5 | $K_1$ | $K_2/K_1$ | Observed |
|---|---|---|---|---|---|---|---|---|---|---|---|
| 10, 1 | 2.1 | (0.01, 0.01, 0.1) | Y | N | N | Y | Y | - | 0.1 | 1.0 | No QSSA |
|  |  | (0.1, 0.1, 1.0) | Y | N | N | Y | Y | - |  |  |  |
|  |  | (1.0, 1.0, 10.0) | Y | N | N | Y | Y | - |  |  |  |
|  | 0.6 | (0.01, 1.0, 0.1) | Y | N | Y | Y | Y | - | 0.1 | 100.0 | QSSA |
|  |  | (0.1, 10.0, 1.0) | Y | N | Y | Y | Y | - |  |  |  |
|  |  | (1.0, 100.0, 10.0) | Y | N | Y | Y | Y | - |  |  |  |
| 1, 1 | 0.9 | (0.1, 0.08, 0.1) | N | N | N | N | N | - | 1.0 | 0.8 | No QSSA |
|  |  | (1.0, 0.8, 1.0) | N | N | N | N | N | - |  |  |  |
|  |  | (10.0, 8.0, 10.0) | N | N | N | N | N | - |  |  |  |
|  | 0.9 | (0.001, 1.0, 0.1) | N | N | Y | Y | Y | - | 0.01 | 1000.0 | QSSA |
|  |  | (0.01, 10.0, 1.0) | N | N | Y | Y | Y | - |  |  |  |
|  |  | (0.1, 100.0, 10.0) | N | N | Y | Y | Y | - |  |  |  |
| 1, 10 | 0.9 | (0.01, 0.08, 0.1) | N | Y | Y | N | N | Y | 1.0 | 0.8 | No QSSA |
|  |  | (0.1, 0.8, 1.0) | N | Y | Y | N | N | Y |  |  |  |
|  |  | (1.0, 8.0, 10.0) | N | Y | Y | N | N | Y |  |  |  |
|  | 4.4 | (0.0001, 0.01, 0.1) | N | Y | Y | Y | Y | Y | 0.01 | 10.0 | QSSA |
|  |  | (0.001, 0.1, 1.0) | N | Y | Y | Y | Y | Y |  |  |  |
|  |  | (0.01, 1.0, 10.0) | N | Y | Y | Y | Y | Y |  |  |  |

Table 2. Characteristics of sets chosen in ref. [21] with varying ratios of $s_0/e_0$ and $K_2/K_1$ and at several $K_1$ values. The predictions are based on (22) and (29). The sets are defined in terms of ($k_1$, $k_{-1}$, $k_2$) given in ref. [21].

| $s_0, e_0$ | *Set | Conditions 2(a) | 2(b) | 2(c) | 2(d) | 3 | 5 | $K_1$ | $K_2/K_1$ | Expected |
|---|---|---|---|---|---|---|---|---|---|---|
| 10, 1 | A | Y | N | Y | Y | Y | - | 0.001 | 10.0 | QSSA |
| | B | Y | N | Y | Y | Y | - | 0.1 | 1000.0 | QSSA |
| | C | Y | N | Y | Y | Y | - | 0.01 | 0.1 | |
| | D | Y | N | Y | Y | Y | - | 10.0 | 100.0 | |
| | E | Y | N | N | N | Y | - | 100.0 | 0.001 | |
| | F | Y | N | N | N | Y | - | 1000.0 | 0.01 | |
| | G | Y | N | N | N | Y | - | 1.0 | 1.0 | |
| 1, 1 | A | N | N | Y | Y | Y | - | 0.001 | 10.0 | QSSA |
| | B | N | N | Y | Y | Y | - | 0.1 | 1000.0 | QSSA |
| | C | N | N | Y | Y | Y | - | 0.01 | 0.1 | |
| | D | N | N | Y | Y | Y | - | 10.0 | 100.0 | |
| | E | N | N | N | N | N | - | 100.0 | 0.001 | |
| | F | N | N | N | N | N | - | 1000.0 | 0.01 | |
| | G | N | N | N | N | N | - | 1.0 | 1.0 | |
| 1, 10 | A | N | Y | Y | Y | Y | Y | 0.01 | 1.0 | |
| | B | N | Y | Y | Y | Y | Y | 1.0 | 100.0 | |
| | C | N | Y | Y | Y | Y | Y | 0.1 | 0.01 | |
| | D | N | Y | Y | Y | Y | Y | 100.0 | 10.0 | |
| | E | N | Y | N | N | N | Y | 1000.0 | 0.0001 | |
| | F | N | Y | N | N | N | Y | 10000.0 | 0.001 | |
| | G | N | Y | N | N | N | Y | 10.0 | 0.1 | No QSSA |

*See ref. [21]

**Figure captions**

Figure 1. Plots of the scaled complex concentration $\gamma$ as a function of scaled time $\tau$ at $s_0/e_0 = 10$. Note that only a small value of $K_1$ does not guarantee QSSA and that a small $\gamma_c$ is not mandatory.

Figure 2. Plots of the scaled complex concentration $\gamma$ as a function of scaled time $\tau$ at $s_0/e_0 = 1$. Note that a small $\tau_c$ is never an indicator of QSSA.

Figure 3. Plots of the scaled complex concentration $\gamma$ as a function of scaled time $\tau$ at $s_0/e_0 = 1/10$. Note again that a small $\tau_c$ cannot ensure QSSA. Also, here $\gamma_c$ is small, but $\tau_c$ is large. This is opposite to the findings of Figure 1.

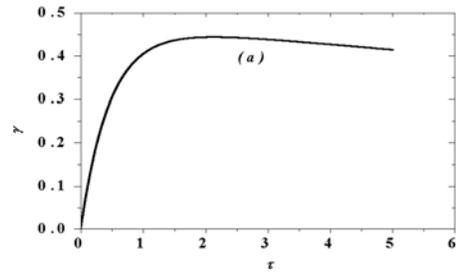

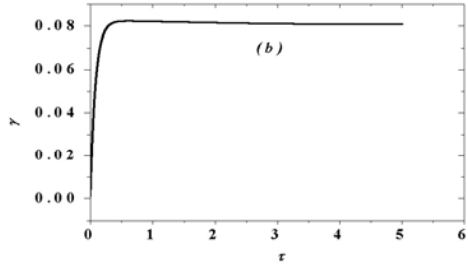

Fig 1

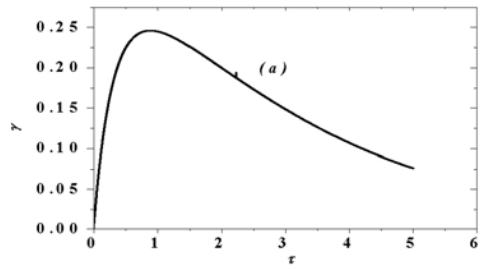

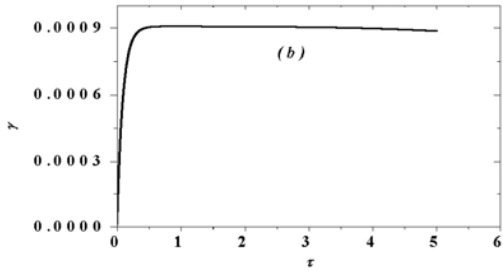

Fig 2

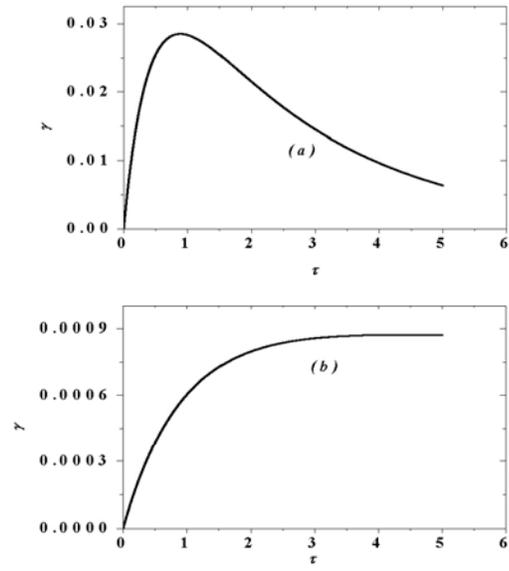

Fig 3